\documentclass[aps,prd,preprint,preprintnumbers,unsortedaddress,superscriptaddress,showpacs,nofootinbib]{revtex4-1}

\usepackage{graphicx,color}
\usepackage{relsize}
\usepackage{slashed}
\usepackage{color}
\usepackage{ulem}
\usepackage{tabu}
\usepackage{bm}

\usepackage{amsmath}
\usepackage{amssymb}
\usepackage{amsthm}
\usepackage{mathrsfs}
\usepackage{graphicx}
\usepackage{epstopdf}
\usepackage{fancyhdr}
\usepackage{array}
\usepackage[all]{xy}
\usepackage{eufrak}
\usepackage{euscript}
\usepackage{enumerate}
\usepackage{slashed}
\usepackage{hyperref}
\usepackage{subfigure} 
\usepackage{epstopdf} 

\hypersetup{pdftex,colorlinks=true,linkcolor=blue,citecolor=blue,menucolor=black,urlcolor=blue,filecolor=blue}

\newcommand{\minv}{M_{\rm inv}}

\newcommand{\mev}{{\rm MeV}}

\renewcommand\sout{\bgroup \color[rgb]{1,0,0} \ULdepth=-.5ex \ULset}

\begin{document}

\title{The $\boldsymbol{\chi_{cJ}}$ decay to $\boldsymbol{\phi K^* \bar K, \phi h_1(1380)}$ testing the nature of axial vector meson resonances}

\author{Sheng-Juan Jiang}
\affiliation{Department of Physics, Guangxi Normal University, Guilin 541004, China}
\affiliation{Guangxi Key Laboratory of Nuclear Physics and Technology, Guangxi Normal University, Guilin 541004, China}

\author{S.~Sakai}
\email{shsakai@itp.ac.cn}
\affiliation{Institute of Theoretical Physics, CAS, Zhong Guan Cun East Street 55 100190 Beijing, China}

\author{Wei-Hong Liang}
\email{liangwh@gxnu.edu.cn}
\affiliation{Department of Physics, Guangxi Normal University, Guilin 541004, China}
\affiliation{Guangxi Key Laboratory of Nuclear Physics and Technology, Guangxi Normal University, Guilin 541004, China}

\author{E.~Oset}
\email{oset@ific.uv.es}
\affiliation{Departamento de F\'{i}sica Te\'{o}rica and IFIC, Centro Mixto Universidad de Valencia - CSIC,
Institutos de Investigaci\'{o}n de Paterna, Aptdo. 22085, 46071 Valencia, Spain}
\affiliation{Department of Physics, Guangxi Normal University, Guilin 541004, China}

\begin{abstract}
 We perform a theoretical study of the $\chi_{cJ} \to \phi K^* \bar K \to \phi K\pi \bar K$ reaction taking into account the $K^* \bar K$ final state interaction,
which in the chiral unitary approach is responsible, together with its coupled channels, for the formation of the low lying axial vector mesons,
in this case the $h_1(1380)$ given the selection of quantum numbers.
Based on this picture we can easily explain why in the $\chi_{c0}$ decay the $h_1(1380)$ resonance is not produced,
and, in the case of $\chi_{c1}$ and $\chi_{c2}$ decay, why a dip in the $K^+ \pi^0 K^-$ mass distribution appears in the 1550-1600 MeV region,
that in our picture comes from a destructive interference between the tree level mechanism and the rescattering that generates the $h_1(1380)$ state.
Such a dip is not reproduced in pictures where the nominal $h_1(1380)$ signal is added incoherently to a background, which provides support to the picture where the resonance appears from rescattering of vector-pseudoscalar components.
\end{abstract}



\maketitle


The BESIII collaboration measured the $\chi_{cJ} \to \phi K^*(892) \bar K$ decay and found a clean signal around 1412 MeV and width 84 MeV
that was associated to the $h_1(1380)$ production \cite{BES}.
As usual in experimental analysis, a Breit-Wigner shaped resonance was added incoherently to a background in the analysis
and a fair reproduction of the data was found, except in the region 1550-1600 MeV where the data fall below the fitted results, showing a pronounced dip.
In the present note we provide an explanation for this dip which is directly tied to the microscopic production process
and the nature of the $h_1(1380)$ as a dynamically generated resonance.

The low lying axial vector meson resonances are fairly well described in a molecular picture from the vector-pseudoscalar interactions
in $s$-wave using the interaction provided by chiral Lagrangians \cite{9:Birse}
and a proper unitary procedure in vector-pseudoscalar coupled channels \cite{5:Lutz,6:Roca,7:GengLS}.
It is then clear that within this picture the $\chi_{cJ} \to \phi h_1(1380)$ must proceed via the production $\chi_{cJ} \to \phi VP$
(with $V$ and $P$ denoting vector and pseudoscalar mesons, respectively), and a posterior interaction of $VP$ that will generate the resonance.

The first step in our analysis is to provide a picture for $\chi_{cJ} \to VVP$, where one of the vectors in particular will be the $\phi$.
Since we are interested only in the shape of the final $K^* \bar K$ mass distribution, we can ignore the strength of this vertex,
but we must relate the different possible trios with two vectors and one pseudoscalar.
For this we are guided by theory and experiment.
From the theoretical point of view we assume that $\chi_{cJ}$ is a SU(3) singlet since $\chi_{cJ}$, made of $c\bar c$, does not contain light quarks.
Then we have a primary structure which is the trace $\langle VVP \rangle$ of the vector and pseudoscalar SU(3) matrices
\begin{equation}\label{eq:Vmatrix}
V = \left(
           \begin{array}{ccc}
             \frac{1}{\sqrt{2}}\rho^0 + \frac{1}{\sqrt{2}}\omega  & \rho^+ & K^{*+} \\
             \rho^- & -\frac{1}{\sqrt{2}}\rho^0 + \frac{1}{\sqrt{2}}\omega  & K^{*0} \\
            K^{*-} & \bar{K}^{*0} & \phi \\
           \end{array}
         \right),
\end{equation}
\begin{equation}\label{eq:phimatrix}
P = \left(
           \begin{array}{ccc}
             \frac{1}{\sqrt{2}}\pi^0 + \frac{1}{\sqrt{3}}\eta + \frac{1}{\sqrt{6}}\eta' & \pi^+ & K^+ \\
             \pi^- & -\frac{1}{\sqrt{2}}\pi^0 + \frac{1}{\sqrt{3}}\eta + \frac{1}{\sqrt{6}}\eta' & K^0 \\
            K^- & \bar{K}^0 & -\frac{1}{\sqrt{3}}\eta + \sqrt{\frac{2}{3}}\eta' \\
           \end{array}
         \right),
\end{equation}
where in $P$ we have considered to the $\eta$-$\eta'$ mixing of Ref. \cite{49:Bramon}.

In the study of the $\chi_{c1} \to \eta \pi^+ \pi^-$ reaction \cite{besKornicer},
it was shown that the structure $\langle PPP \rangle$ for the vertex was favored by the experiment
and the other possible structures $\langle PP \rangle \langle P \rangle$
and $\langle P \rangle^3$ were clearly rejected by experiment \cite{46:Liang:chic1,48:Vinicius:etac}.
In analogy to this, and prior to that work, it was found in Refs. \cite{44:UGMOller,45:Roca}
that in the $J/\psi \to \phi \pi^+ \pi^-, \omega \pi^+ \pi^-$ reactions the most important structure was again $\langle VPP \rangle$,
with a small component of $\langle V \rangle \langle PP \rangle$.  
This structure was again used in Ref. \cite{LiangSakai}
to study the $J/\psi \to \eta' h_1, \eta h_1, \pi^0 b_1$ reactions in connection with the BESIII experiment \cite{etaprime},
where a fair agreement with experimental results was obtained using the same interaction as in Refs. \cite{44:UGMOller,45:Roca}.
In analogy to this, we propose now the structure
\begin{equation}\label{eq:H}
  H = {\mathcal{C} \langle VVP \rangle},
\end{equation}
which should be dominant, where $\mathcal C$ is an arbitrary constant.
A possible mixture of $\langle VV \rangle \langle P \rangle$ could change a bit the production rates,
but we are only interested in the shape of the $K^* \bar K$ mass distribution which is not affected by this small admixture.

Equation \eqref{eq:H}, using the $P$ and $V$ matrices of Eqs. \eqref{eq:phimatrix} and \eqref{eq:Vmatrix}, gives the structure
\begin{equation}\label{eq:H1}
  H= {\mathcal C} \left[ K^{*+}\phi K^- + K^{*0} \phi \bar K^0 + \phi K^{*-} K^+ + \phi \bar K^{*0} K^0
  + \phi \, \phi \left( \frac{-\eta}{\sqrt{3}} + \sqrt{\frac{2}{3}} \eta' \right)\right],
\end{equation}
which, ignoring the order and singling out the $\phi$ field, gives
\begin{equation}\label{eq:H2}
  H= {\mathcal C} \,\phi \left[ K^{*+} K^- + K^{*0} \bar K^0 + K^{*-} K^+ + \bar K^{*0} K^0 + \phi \left( \frac{-\eta}{\sqrt{3}} + \sqrt{\frac{2}{3}} \eta' \right)\right].
\end{equation}
Given the large mass of $\eta'$, we neglect it in our study, as was also done in Ref. \cite{6:Roca}.

The structure with kaons that we have in Eq. \eqref{eq:H2} corresponds to the $\bar K^* K$ combination of isospin $I=0$ and $C$-parity $C=-$,
with our convention ($K^- =- | 1/2, -1/2\rangle, K^{*-}=-| 1/2, -1/2\rangle, C K^{*+}= -K^{*-}$),
\begin{equation}\label{eq:combKsK0}
\frac{1}{\sqrt{2}}\left(|\bar K^* K\rangle_{I=0}-|K^* \bar K\rangle_{I=0} \right) =\frac{1}{2} (\bar K^{*0} K^0+ K^{*-} K^+ + K^{*+} K^- + K^{*0} \bar K^0 ).
\end{equation}
This convention is the same one used in Ref. \cite{6:Roca} and we can then take the coupling of the $h_1(1380)$ to the different channels from Ref. \cite{6:Roca}.
The $\eta \phi$ term has also $I=0, C=-$, as it should be to match, together with the other $\phi$, the $\chi_{cJ}$ states.

\begin{figure}[h!]
\begin{center}
\includegraphics[scale=0.6]{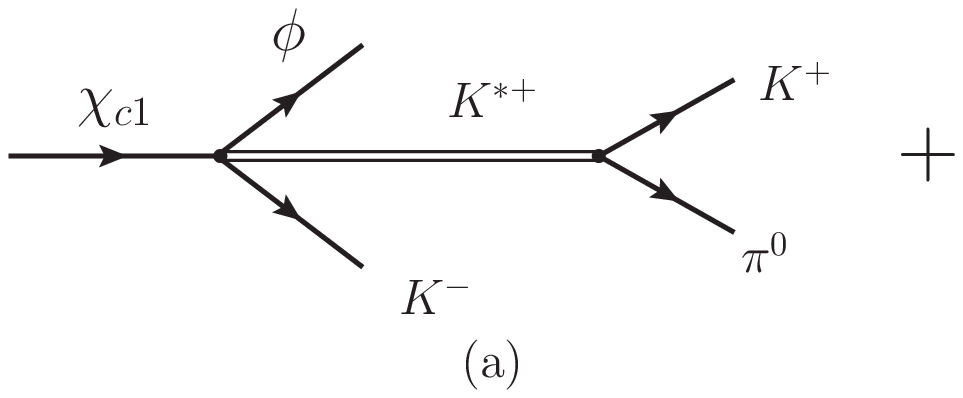}
\includegraphics[scale=0.6]{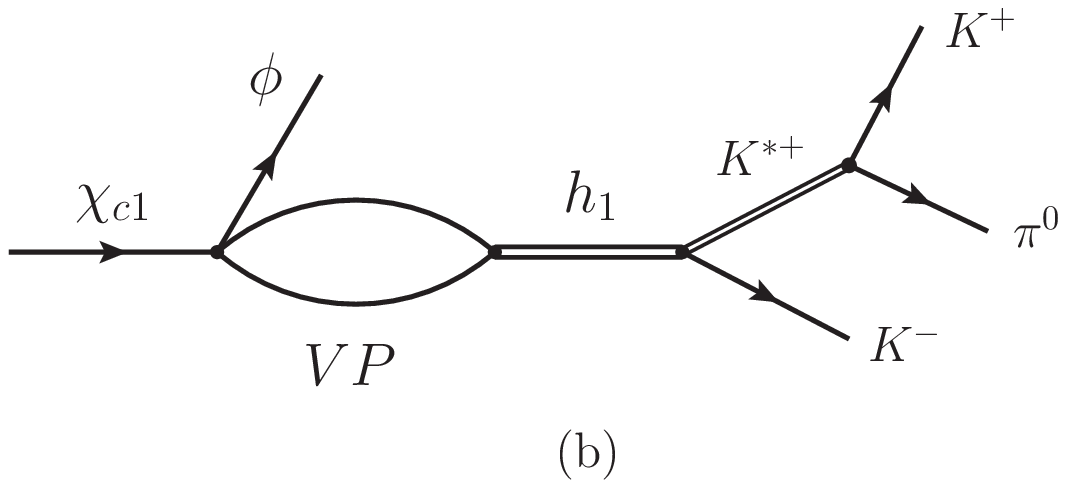}
\end{center}
\vspace{-0.7cm}
\caption{Mechanisms for $\chi_{c1} \to \phi K^{*+} K^- \to \phi K^+ \pi^0 K^-$ reaction,
(a) tree level; (b) rescattering of $VP (K^* \bar K, \bar K^* K, \phi \eta)$ producing the $h_1(1380)$ resonance
that decays into $K^{*+} K^-$ and later in $K^+ \pi^0 K^-$.}
\label{Fig:1}
\end{figure}

Next we must look at the spin-parity structure of the vertices for the different $\chi_{cJ}$ states.
When doing this we must take into account the order in which the fields appear in Eq.~\eqref{eq:H1}:
\begin{enumerate}
\item[1)] $\chi_{c0} \to \phi K^* \bar K$ ~~~[$\chi_{c0}: I^G (J^{PC}) = 0^+ (0^{+\,+})$]\\
Since in the final state we have $V(1^-)\, V(1^-)\, P(0^-)$, we need a $p$-wave to conserve parity.
This eliminates structures like $\vec \epsilon_\phi \cdot \vec \epsilon_{K^*}$ or $(\vec \epsilon \cdot \vec p_K) (\vec \epsilon_{K^*} \cdot \vec p_\phi)$.
The structure must be of the type
\begin{equation}\label{eq:new2}
 (\vec \epsilon_\phi \times \vec \epsilon_{K^*} )\cdot \vec p_i,
\end{equation}
with $\vec p_i$ some of the final momenta.
This is indeed the operator used in the $dd\to \eta ^4 {\rm He}$ reaction \cite{hanhart,Ikeno,liangXie}, which has the same quantum numbers.
With these structures, if one selects for instance the $K^{*+} \phi K^-$ final channel, as shown diagrammatically in Fig. \ref{Fig:1}(a),
one will get some contribution to the decay in this channel, as observed experimentally.
Yet, if we wish to produce the $h_1(1380)$ resonance,
we will have to consider the diagram of  Fig. \ref{Fig:1}(b) and sum coherently in the loop over the states of Eq.~\eqref{eq:H1}.
For this we must take care about the order in which the states appear in Eq.~\eqref{eq:H1}.
Take first the term with momentum $\vec p_K$ in Eq.~\eqref{eq:new2}.
We will have the combination
\begin{equation}\label{eq:newstru1}
 2 (\vec \epsilon_{K^*} \times \vec \epsilon_{\phi} )\cdot \vec p_K + 2 (\vec \epsilon_\phi \times \vec \epsilon_{K^*} )\cdot \vec p_K
 -\frac{1}{\sqrt{3}}(\vec\epsilon_{\phi_1} \times \vec\epsilon_{\phi_2}) \cdot \vec p_\eta
 -\frac{1}{\sqrt{3}}(\vec\epsilon_{\phi_2} \times \vec\epsilon_{\phi_1}) \cdot \vec p_\eta=0.
\end{equation}

Next we would take the momentum of the two vectors and by symmetry we will have again
\begin{eqnarray}\label{eq:newstru2}
    && 2 (\vec \epsilon_{K^*} \times \vec \epsilon_{\phi} )\cdot \vec p_\phi +2 (\vec \epsilon_\phi \times \vec \epsilon_{K^*} )\cdot \vec p_{K^*}
  +2(\vec \epsilon_{K^*} \times \vec \epsilon_{\phi} )\cdot \vec p_{K^*} +2 (\vec \epsilon_\phi \times \vec \epsilon_{K^*} )\cdot \vec p_\phi \nonumber \\
    &-&\frac{1}{\sqrt{3}}\left[
    (\vec\epsilon_{\phi_1} \times \vec\epsilon_{\phi_2}) \cdot \vec p_{\phi_2}
    +(\vec\epsilon_{\phi_1} \times \vec\epsilon_{\phi_2}) \cdot \vec p_{\phi_1}
    +(\vec\epsilon_{\phi_2} \times \vec\epsilon_{\phi_1}) \cdot \vec p_{\phi_1}
    +(\vec\epsilon_{\phi_2} \times \vec\epsilon_{\phi_1}) \cdot \vec p_{\phi_2} \right]\nonumber \\
    &=&0.
 \end{eqnarray}
And we see that in the coherent sum the terms cancel and there is no $h_1(1380)$ production.
This is the first output of our approach, since the assumed nature of the $h_1(1380)$ has as a consequence
that the $h_1(1380)$ is not produced in the $\chi_{c0} \to \phi K^* \bar K$ reaction.
This is corroborated by the experimental findings of Ref. \cite{BES}.

\item[2)] $\chi_{c1} \to \phi K^* \bar K$ ~~~[$\chi_{c1}: I^G (J^{PC}) = 0^+ (1^{+\,+})$]\\
Once again we need a $p$-wave and hence a momentum of the final particles.
The argumentation is easy in the rest frame of $K^* \bar K (\bar K^* K)$.
We will have a structure of the type
\begin{equation}\label{eq:new4}
  (\vec \epsilon_{\chi_{c1}} \cdot \vec p_i)(\vec \epsilon_\phi \cdot \vec \epsilon_{K^*}),
\end{equation}
with $p_i$ any of the final momenta, or any cyclical permutation of this form
(the $\phi \phi \eta$ term can be equally considered with $\epsilon_\phi \to \epsilon_{\phi_1}$, $\epsilon_{K^*} \to \epsilon_{\phi_2}$,
plus keeping the symmetry of 1,2 for the two identical $\phi$ mesons).
Take now the channel $K^{*+} \phi K^-$ of Eq.~\eqref{eq:H1}.
Considering the symmetry of the two vectors we will have the combinations for the tree level of Fig. \ref{Fig:1}(a),
\begin{equation}\label{eq:struchic1}
 {\mathcal{C}} (\vec \epsilon_{\chi_{c1}} \cdot \vec p_K) \; (\vec \epsilon_\phi \cdot \vec \epsilon_{K^*}),
\end{equation}
\begin{equation}
   {\mathcal{C'}} \left[(\vec \epsilon_{\chi_{c1}} \cdot \vec p_{K^*}) + (\vec \epsilon_{\chi_{c1}} \cdot \vec p_\phi)\right]
 \, (\vec \epsilon_\phi \cdot \vec \epsilon_{K^*} ).
\end{equation}
The terms that go with $\vec p_K$ or $\vec p_{K^*}$ involve $p$-wave and in the loops of Fig. \ref{Fig:1}(b) they will vanish. 

Hence, in the coherent sum of Fig. \ref{Fig:1}(b) for the loop of $VP$ we will get
\begin{equation}\label{eq:new6}
  (\vec \epsilon_\phi \cdot \vec \epsilon_{K^*} ) \, (\vec \epsilon_{\chi_{c1}} \cdot \vec p_\phi),
\end{equation}
which will be multiplied by the $G$ function and the $h_1$ amplitude later.
This term will then interfere with the $s$-wave term of the tree level which has the same structure.
Other terms in the tree level, which involve $p$-wave in $\vec p_K$, would not interfere with the loop term and would go into a background.

The argument can be extended to any of the cyclical combinations of Eq.~\eqref{eq:new4} with the same results,
factorizing the same term involving $\vec p_\phi$ in the tree level and the loop contribution.
Since we have an arbitrary normalization at the end, the whole discussion can be done with just the structure of Eq.~\eqref{eq:new6}.

\item[3)] $\chi_{c2} \to \phi K^* \bar K$ ~~~[$\chi_{c2}: I^G (J^{PC}) = 0^+ (2^{+\,+})$]\\
Given the symmetry between the two vectors and what was found in points 1) and 2),
it is clear that we should now combine the $\phi$ and $K^*$ spin to $J=2$ to avoid the cancellations found in point 1)
where $J=1$ ( $\vec \epsilon_\phi \times \vec \epsilon_{K^*}$ combination).
The spin 2 of the $\chi_{c2}$ can be combined with the $\vec p_\phi$
to give another tensor of rank two to be contracted with the $J=2$ object constructed with the two vectors.
We would have remaining terms in the tree level and the loops involving $\vec p_\phi$ that would produce the $h_1(1380)$ resonance
and some interference between them.
We do not elaborate further since the calculations in what follow are only done for $\chi_{c1}$ decay.
Experimentally one finds that the signal of $h_1(1380)$ is clearly seen in the $\chi_{c1}$ and $\chi_{c2}$ decays \cite{BES}.
\end{enumerate}

Next we must consider that in the experiment the $K^*$ is seen as a $K\pi$ state.
If we want to have $K^+ K^- \pi^0$, as experimentally measured, we can have
\begin{equation*}
 \chi_{c1} \to \phi K^{*+} K^-,~~ K^{*+} \to K^+ \pi^0,
\end{equation*}
or
\begin{equation*}
 \chi_{c1} \to \phi K^{*-} K^+,~~ K^{*-} \to K^- \pi^0.
\end{equation*}
In these processes the $K^+$ is in $p$-wave in the first case and the $K^-$ is in $p$-wave in the second case.
There is no interference upon angle integrations between the two mechanisms and their contributions would be the same.
Since we are concerned only about the shape of the distributions,
we consider only the first mechanism that we depict in  Fig. \ref{Fig:1}.

The amplitude corresponding to the mechanisms of  Fig. \ref{Fig:1} is given by
\begin{eqnarray}\label{eq:t}
t&=&A \;h_{K^{*+}K^-} \, D_{K^*}(\minv (K^+ \pi^0))\, (\vec \epsilon_{\chi_{c1}} \cdot \vec p_\phi) \; (\vec \epsilon_\phi \cdot \vec \epsilon_{K^*}) \;
\vec \epsilon_{K^*} \cdot (\vec p_K -\vec p_\pi)\nonumber \\
&& + A\; h_{K^{*} \bar K} \, G_{K^* \bar K}(\minv (K^{*+} K^-))\, (\vec \epsilon_{\chi_{c1}} \cdot \vec p_\phi) \;
2\, g_{h_1, K^{*} \bar K} \; g_{h_1, K^{*+} K^-}
\;D_{h_1}(\minv (K^{*+}K^-)) \nonumber \\
&& ~~\times  D_{K^*}(\minv (K^+ \pi^0))\;(\vec \epsilon_\phi \cdot \vec \epsilon_{K^*})\; (\vec \epsilon_{K^*} \cdot \vec \epsilon_{h_1})\;
(\vec \epsilon_{h_1} \cdot \vec \epsilon_{K^*})\; \vec \epsilon_{K^*} \cdot (\vec p_K - \vec p_\pi)\nonumber \\
&&+A\; 2 h_{\phi \eta}\, G_{\phi \eta}(\minv (K^{*+}K^-))\; (\vec \epsilon_{\chi_{c1}} \cdot \vec p_\phi)\; g_{h_1, \phi \eta}\; g_{h_1, K^{*+}K^-} \;
D_{h_1}(\minv (K^{*+}K^-))  \nonumber \\
&&~~ \times  D_{K^*}(\minv (K^+ \pi^0)) \;(\vec \epsilon_\phi \cdot \vec \epsilon_{K^*})\; (\vec \epsilon_{K^*} \cdot \vec \epsilon_{h_1})\;
(\vec \epsilon_{h_1} \cdot \vec \epsilon_{K^*})\; \vec \epsilon_{K^*} \cdot (\vec p_K - \vec p_\pi),
\end{eqnarray}
where $A$ is a constant that contains $\mathcal C$ of Eq.~\eqref{eq:H2} and a factor from the $K^* \to K\pi$ coupling,
and the factor $\vec p_K - \vec p_\pi$ must be taken in the $K^*$ rest frame.
In addition, $h_{K^{*+} K^-}$ and $h_{\phi \eta}$ are the weights of these states in the $H$ combination of Eq.~\eqref{eq:H2}
and $h_{K^* \bar K}$ the weight of the whole $K^* \bar K, \bar K^* K$ combination in Eq.~\eqref{eq:H2},
\begin{equation}\label{eq:hfactor}
  h_{K^{*} \bar K}=1, ~~~~~~h_{K^{*+} K^-}=1, ~~~~~~ h_{\phi \eta}= - \frac{1}{\sqrt{3}}.
\end{equation}
The factor $2$ in front of $h_{\phi \eta}$ in Eq.~\eqref{eq:t} stems from the identity of $\phi \phi$ in the Hamiltonian.
The coupling $g_{h_1, K^{*} \bar K}$ stands for the coupling of $h_1$ to the combination of Eq.~\eqref{eq:combKsK0},
which is the one reported in Ref.~\cite{6:Roca}.
Similarly, $g_{h_1, \phi \eta}$ is also taken from Ref.~\cite{6:Roca}.
The factor $2$ in front of $g_{h_1, K^{*} \bar K}$ in Eq.~\eqref{eq:t} is because of the normalization of this state in $H$ of Eq.~\eqref{eq:H2}
compared to the normalization of the $K^* \bar K$ wave function of Ref.~\cite{6:Roca} given in Eq.~\eqref{eq:combKsK0}.
In addition, $g_{h_1, K^{*+}K^-}= \frac{1}{2} g_{h_1, K^{*} \bar K}$ of Ref.~\cite{6:Roca}.
The couplings of $h_1(1380)$ to $K^* \bar K$ and $\phi \eta$ from Ref.~\cite{6:Roca} are
\begin{equation}\label{eq:g}
 g_{h_1, K^{*} \bar K} = 6147+i 183 \; \mev,  ~~~~~ g_{h_1, \phi \eta} = -3311+i 47\; \mev,  ~~~~~ g_{h_1, K^{*+}K^-}= \frac{1}{2} g_{h_1, K^{*} \bar K}.
\end{equation}
The $D_{h_1}$, $D_{K^*}$ propagators of Eq.~\eqref{eq:t} are given by
	 \begin{align}
	  D_{h_1}(\minv (K^{*+}K^-))= \frac{1}{\minv^2 (K^{*+}K^-) -m_{h_1}^2 + i m_{h_1} \Gamma_{h_1}},\label{eq:Dh1}\\
	  D_{K^*}(\minv (K^+ \pi^0))=  \frac{1}{\minv^2 (K^{+}\pi^0) -m_{K^*}^2 + i m_{K^*} \Gamma_{K^*}},\label{eq:DKstar}
	 \end{align}
with the mass $m_{h_1}$ and width $\Gamma_{h_1}$ of $h_1(1380)$ in the PDG \cite{pdg2018},
and $G_i$ are the $VP$ loop functions that we take from Ref.~\cite{6:Roca} using dimensional regularization with the same subtraction constant.
In $G_{K^*\bar{K}}$, the width of $K^*$ is taken into account by means of a convolution using the $K^*$ spectral function.

Summing over the $K^*$ and $h_1$ polarizations in  Eq.~\eqref{eq:t}, we find
\begin{equation}\label{eq:t2}
  t= A\; (\vec \epsilon_{\chi_{c1}} \cdot \vec p_{\phi})\; \vec \epsilon_{\phi} \cdot (\vec p_K - \vec p_\pi) \; D_{K^*}(\minv (K^+ \pi^0)) \; T,
\end{equation}
where $T$ is given by
\begin{eqnarray}\label{eq:T1}
T = h_{K^{*+} K^-} + D_{h_1}(\minv (K^{*+}K^-)) &\left[ \right. & h_{K^{*} \bar K} \; g^2_{h_1, K^{*} \bar K} \; G_{K^* \bar K}(\minv (K^{*+} K^-))  \nonumber \\
&+&  h_{\phi \eta}\, g_{h_1, \phi \eta}\; g_{h_1, K^{*} \bar K}\; G_{\phi \eta}(\minv (K^{*+}K^-)) \left. \right].
\end{eqnarray}
When we sum and average over polarizations in $|t|^2$, we find at the end
\begin{equation}\label{eq:sumt2}
  \overline{\sum} \sum |t|^2 = B\, \vec p^{\,2}_\phi \, \tilde{p}^2_{K^+} \, |D_{K^*}|^2 \; |T|^2,
\end{equation}
where in $B$ we concentrate the different constant factors,
and the double differential mass distribution is given by \cite{51:Pavao}
\begin{equation}\label{eq:ddGama}
\frac{{\rm d}^2 \Gamma}{{\rm d} \minv (K^{*+} K^-)\, {\rm d}\minv (K^+ \pi^0)}= \frac{1}{(2\pi)^5} \, p_\phi \, p_{K^-} \, \tilde{p}_{K^+}
\; \frac{1}{4 M^2_{\chi_{c1}}} \; \overline{\sum} \sum |t|^2,
\end{equation}
where
\begin{eqnarray}
p_\phi &=& \frac{\lambda^{1/2} (M^2_{\chi_{c1}}, m^2_\phi, \minv^2 (K^{*+} K^-))}{2\, M_{\chi_{c1}}}, \label{eq:pphi}\\
 p_{K^-} &=& \frac{\lambda^{1/2} (\minv^2 (K^{*+} K^-), m^2_{K^-}, \minv^2 (K^+ \pi^0))}{2\, \minv (K^{*+} K^-)}, \label{eq:pK2}\\
 \tilde{p}_{K^+} &=& \frac{\lambda^{1/2} (\minv^2 (K^+ \pi^0), m^2_{K^+}, m^2_{\pi^0})}{2\, \minv^2 (K^+ \pi^0)}. \label{eq:pK1}
\end{eqnarray}

We integrate the differential width of Eq.~\eqref{eq:ddGama} over $\minv (K^+ \pi^0)$
and compare the $\frac{{\rm d} \Gamma}{{\rm d} \minv (K^{*+} K^-)}$ distribution with experiment.

For the $\chi_{c2} \to \phi K^+ \pi^0 K^-$ we would get the same formulas up to a possible different constant $B$
and the different mass of the $\chi_{c2}$.
We have seen that the shape of the $K^+ K^- \pi^0$ mass distribution for the $\chi_{c2}$ decay is practically the same as for the $\chi_{c1}$ decay
in the range that we are interested in.
Actually this is also the case for the BESIII experiment \cite{BES}.
Finally, we do not evaluate the mass distribution for the $\chi_{c1} \to K_S^0 K^\pm \pi^\mp$ decay,
but based on the possible $K^* \bar K$, $\bar K^* K$ modes leading to this distribution we get a rate twice as large as for the $K^+ K^- \pi^0$ decay,
as clearly seen in the experiment,
and the same shape as the one calculated.
In view of these findings, we can compare the results that we obtain, up to an arbitrary normalization, with those of Ref.~\cite{BES} for the sum of all these modes,
which is done to gain statistics.

\begin{figure}[bt]
\begin{center}
\includegraphics[scale=0.7]{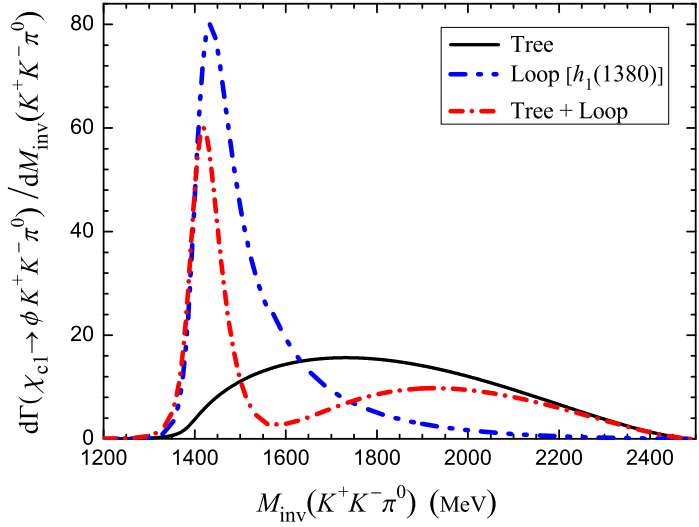}
\end{center}
\vspace{-0.7cm}
\caption{Contributions of the terms of Eq. \eqref{eq:T1} to $\frac{{\rm d} \Gamma}{{\rm d} \minv (K^{*+} K^-)}$ seen in $\chi_{c1} \to \phi K^{+} K^- \pi^0$ decay.}
\label{Fig:2}
\end{figure}

In Fig.~\ref{Fig:2} we can see the contribution to $\frac{{\rm d} \Gamma}{{\rm d} \minv (K^{*+} K^-)}$
from the first term of $T$ in Eq.~\eqref{eq:T1} (tree level),
the second term that contains the $h_1(1380)$ propagator (``Loop'' in the figure) and the coherent sum.
The interesting finding is that the tree level, which by itself could be considered a background and contributes basically according to phase space,
interferes destructively with the $h_1(1380)$ signal and the resulting shape is quite different from the one of the $h_1(1380)$ itself,
which has a much broader shape.
It is interesting to see that our contribution of the $h_1(1380)$ alone has basically the same shape as the one of Ref.~\cite{BES} in Fig. 7 of that work.
In addition, in Ref.~\cite{BES} a background, and the $\phi(1680)$ and $\phi(1850)$ contributions are added incoherently.
The interference between the tree level mechanism and the $h_1(1380)$ contribution is then missed in that analysis.
It is clear that our approach will show a dip in the region of 1550-1600 MeV of the $K^+ K^- \pi^0$ invariant mass.

\begin{figure}[bt]
\begin{center}
\includegraphics[scale=0.7]{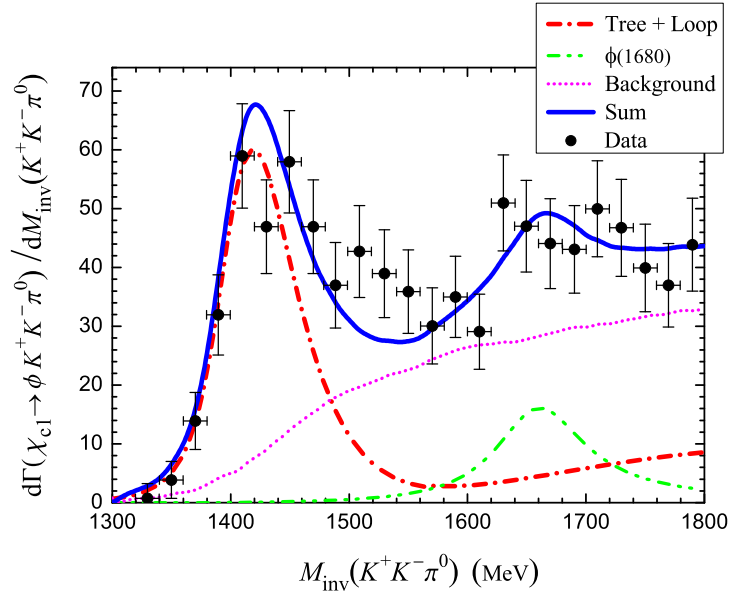}
\end{center}
\vspace{-0.7cm}
\caption{Comparison of our results with the data of Ref.~\cite{BES}.}
\label{Fig:3}
\end{figure}

In order to compare the mass distribution $\frac{{\rm d} \Gamma}{{\rm d} M_{\rm inv}(K^{*+}K^-)}$ of Ref.~\cite{BES},
we also add a background and the $\phi(1680)$ contribution,
since we are not interested in the region where the $\phi(1850)$ can contribute.
We modify minimally the input of Ref.~\cite{BES},
but some different normalization is needed in view of the interference that we have mentioned.
Our signal for the $h_1(1380)$ of the peak is multiplied by 1.2 and the background by 1.3 while,
we keep the same strength at the peak of the $\phi(1680)$ as in Ref.~\cite{BES}.
The results are shown in Fig.~\ref{Fig:3}.
There is also another small difference in the $\phi(1680)$ contribution since its decay into $K^* \bar K$ must proceed in $p$-wave
and hence we have a contribution of the type
\begin{equation}\label{eq:phi1680}
  D\, \left| \frac{\tilde{p}_{K^*}}{\minv^2(K^* \bar K) -M^2_{\phi(1680)}+i M_{\phi(1680)} \Gamma_{\phi(1680)}} \right|^2,
\end{equation}
where $D$ is a constant, and
\begin{eqnarray}
\tilde{p}_{K^*} &=& \frac{\lambda^{1/2} (\minv^2 (K^{*} \bar K), m^2_{K^*}, m^2_K)}{2\, \minv (K^{*} \bar K)}, \label{eq:pt1}\\
 \Gamma_{\phi(1680)} &=& \Gamma^{(0)}_{\phi(1680)}\; \frac{\tilde{p}^3_{K^*}}{\tilde{p}^3_{K^*, {\rm on}}}, \label{eq:pt2}\\
 \tilde{p}_{K^*, {\rm on}} &=& \frac{\lambda^{1/2} (M^2_{\phi(1680)}, m^2_{K^*}, m^2_K)}{2\, M_{\phi(1680)}}, \label{eq:pt3}
\end{eqnarray}
with $M_{\phi(1680)}$ and $\Gamma^{(0)}_{\phi(1680)}$ being the mass and width of $\phi(1680)$.

What we see in Fig.~\ref{Fig:3} is that our approach produces naturally a clear dip in the region of 1550-1600 MeV,
while the fit of Ref.~\cite{BES} gives a distribution that is above the data in that region.
It is clear that the experimental fit will give a larger strength in that region than our approach because they do not have the interference of the background with the resonance that we have shown in our approach.
Certainly a different fit to the data could have been done putting a tree level and the signal of the $h_1(1380)$ and letting them interfere,
however, in that fit, the strength of the tree level and its sign would be uncorrelated.
In our approach the relative strength and sign are given once we assume that the $h_1(1380)$ is generated from the interaction between pseudoscalar and vector.
This is why the dip which we predict for this distribution is tied to the nature of the $h_1(1380)$ as a dynamically generated resonance,
and the fact that this feature is present in the experiment provides a great support for that picture of the $h_1(1380)$,
and by analogy other axial vector meson resonances as dynamically generated from the vector-pseudoscalar interaction.

In summary, based on the picture that the $h_1(1380)$ is a dynamically generated resonance formed from the interaction of vector-pseudoscalar pairs,
mostly $K^*\bar K -c.c.$ and $\phi \eta$,
we have carried out a study of the $\chi_{cJ} \to \phi K^* \bar K \to \phi K^+ \pi^0 K^-$ reaction
and related charge channels and have obtained a fair reproduction of the shape of the experimental data.
Due to the fact that in this picture the $h_1(1380)$ is generated from the $s$-wave $VP$ interaction,
we could justify why no $h_1(1380)$ signal was found in the $\chi_{c0}$ decay,
while the signal appeared both in the $\chi_{c1}$ and $\chi_{c2}$ decays.
Another remarkable feature of the study was that we could determine the relative strength between the tree level contribution
to the process and the one that contains the $h_1(1380)$ production, and we found a destructive interference between the two processes
that significantly distorts the $h_1(1380)$ signal with respect to the Breit-Wigner shape and produces a dip
in the $K^+ \pi^0 K^-$ mass distribution around the region of 1550-1600 MeV.
This dip is present in the experiment and not reproduced in a picture that sums incoherently the $h_1$ Breit-Wigner distribution with a smooth background,
providing a strong support to the molecular picture of the $h_1(1380)$ resonance.

\begin{acknowledgments}
We thank Wen-Biao Yan for useful discussions and suggestions.
This work is partly supported by the National Natural Science Foundation of China (Grants No. 11565007 and No.~11847317).
This work is also partly supported by the Spanish Ministerio de Economia y Competitividad
and European FEDER funds under the contract number FIS2011-28853-C02-01, FIS2011-28853-C02-02, FIS2014-57026-REDT, FIS2014-51948-C2-1-P, and FIS2014-51948-C2-2-P,
and the Generalitat Valenciana in the program Prometeo II-2014/068.
S.~Sakai acknowledges the support by NSFC and DFG through funds provided to the Sino-German CRC110 ``Symmetries and the Emergence of Structure in QCD'' (NSFC Grant No.~11621131001), by the NSFC (Grants No.~11747601 and No.~11835015),
by the CAS Key Research Pro-gram of Frontier Sciences (Grant No.~QYZDB-SSW-SYS013) and by the CAS Key Research Program (Grant No.~XDPB09).
\end{acknowledgments}


  \end{document}